\documentclass[draft]{agujournal2019}
\usepackage{url} 
\usepackage{lineno}
\usepackage[finalnew]{trackchanges} 
\usepackage{soul}
\usepackage{bm}
\usepackage{textcomp, gensymb}

\draftfalse

\journalname{Earth and Space Science}

\begin{document}

%
%


\title{First results from a low cost software defined radio system monitoring VHF equatorial ionospheric scintillations in the Indian sector}

%
%

\authors{
Jishnu N. Thekkeppattu\affil{1}}

\affiliation{1}{International Centre for Radio Astronomy Research (ICRAR), Curtin University, Bentley, WA 6102, Australia}

\correspondingauthor{Jishnu N. Thekkeppattu}{j.thekkeppattu@curtin.edu.au (JNT)}




\begin{keypoints}
\item A low cost system to observe ionospheric scintillations using a software defined radio (SDR) is described.
\item A method to subtract RFI from scintillation measurements using frequency differencing is presented.
\item Analysis of a scintillation patch and a few instances of relatively rare daytime scintillations are also reported.
\end{keypoints}

%
%

%
%


\begin{abstract}
Scintillations of radio frequency signals due to the ionosphere, despite having been studied for decades, is still an active area of research. Of particular interest is the scintillations near the geomagnetic equator, where such scintillations can be strong enough to cause disruptions to satellite communications. In this paper, a low cost system to monitor VHF scintillations is described. The system called the Personal Ionospheric Experiment (PIE) makes use of a widely available software defined radio (SDR) to perform the measurements. The use of an SDR vastly enhances the flexibility, especially in environments with non-negligible radio frequency interference (RFI). A simple technique to effectively subtract out transient RFI commonly found in radio data is detailed, and tested with data collected over several months. The utility of the proposed system is demonstrated with the analysis of a night-time scintillation patch; a few daytime scintillations are also reported. Therefore this paper demonstrates the feasibility of conducting useful low-cost radio science experiments in semi-urban locations with RFI, encouraging similar citizen science initiatives.
\end{abstract}



\section{Introduction}
Fluctuations in the received amplitude and phase of radio signals originating from space, due to irregularities in the ionospheric plasma is commonly referred to as ionospheric scintillation. Ionospheric scintillations are a scientific opportunity and a scientific and engineering challenge. Observations of scintillations can provide valuable insights into the dynamics of the ionosphere, upper atmosphere, magnetosphere and the complex interactions between space weather and Earth systems. However, ionospheric scintillations are also detrimental to radio astronomy observations \cite{2016MNRAS.458.3099V, 2018A&A...615A.179D}, satellite navigation systems \cite{2001RaSc...36..731K, 2006AdSpR..38.2478D, 2007SpWea...5.9003K}, satellite VHF and UHF communication systems \cite{2014SpWea..12..601K} etc. It is therefore imperative that ionosphere and scintillations are monitored.

However, there is a paucity of open access VHF scintillation data to assist in such studies, especially along the geomagnetic equator. Most of the research into ionospheric scintillation is carried out with dedicated scintillation measurement receivers installed in a few research institutes. In studying ionospheric scintillations - which has strong latitudinal and longitudinal variations - arrays of receivers that share a common data format and similar design features could be beneficial. Such an array can also be used to perform cross-correlation analysis to determine certain parameters of the ionospheric plasma \cite{1989JGR....9411959B, 2001GeoRL..28..119B}. Besides, as ionospheric conditions have complex associations with space weather, collection of data over prolonged periods is important.  

With the above goals in mind, an experimental low-cost VHF scintillation monitoring system, and associated data capture and analysis principles are proposed in this paper. The scope of this paper is threefold - describe a proof-of-concept VHF scintillation monitoring system using a software defined radio (SDR), present some early results from it, and provide an open repository of the data collected from it. The experiments are carried out in the 250~MHz downlink band for geostationary/geosynchronous satellites. As these satellites are always visible over the equatorial regions, round-the-clock monitoring of the ionospheric conditions is feasible.

The system described in this paper is called the Personal Ionosphere Experiment (PIE), and is installed at the author's house in the city of Payyanur (latitude 12.1\degree N, longitude 75.2\degree E, magnetic inclination 13.04 \degree N), in the state of Kerala in India. The location is sufficiently close to the geomagnetic equator that strong ionospheric scintillations are expected. Indeed, analysis of data collected over a few months shows promising results, including detections of daytime scintillations. Also central to the data analysis in this paper is a simple but powerful technique to remove impulsive RFI from these data, which plagues radio science experiments. 

This paper is organized as follows. In Sec.\ref{sec:sys_desc}, a description of the system is given, along with details of the acquisition software and the data format. In Sec.\ref{sec:data_rfi}, preliminary inspection of raw data, details of RFI subtraction and RFI subtracted scintillation timeseries are provided. In Sec.\ref{sec:results}, the system is commissioned by carrying out power spectrum analysis of a scintillation patch and comparing the results obtained with literature. In the same section, a few instances of daytime scintillations captured by the system are also reported. The paper is concluded with a discussion on the potential citizen science applications.
\section{System description}
\label{sec:sys_desc}
The original intention behind the system described in this paper was to setup a general purpose radio science experiment to observe meteor reflections in the 88-108~MHz FM band as well as solar radio bursts at the frequencies above 200~MHz. The system is therefore designed to perform unattended continuous data collection and hence the receiver chain is kept simple for easy maintenance and to reduce the number of points of failure. The hardware consists of an antenna with a low noise amplifier, an RTL-SDR Blog V3 SDR attached to an Orange Pi Zero2 single board computer (SBC) which is connected to the residential internet installation. The antenna is similar to a Yagi-Uda antenna \cite{2002aaa..book.....K} consisting of a dipole and a reflector, however no director is used. The antenna is tuned to about 150~MHz and is constructed with RG-58 coaxial cables with the shield and the centre conductor shorted. The frequency of 150~MHz was chosen as a good trade-off between the original requirements of the experiment and antenna dimension limitations. The dipole is immediately followed by a low noise amplifier (LNA) based on INA-02184 MMIC (noise figure$\sim$3dB), modified to be used in a bias-T configuration. The coaxial cable connecting the balanced dipole to the unbalanced LNA is wound on a ferrite core, which acts as a simple choke balun. The antenna, balun and the amplifier are enclosed in plumbing PVC tubing to weather-proof them. The antenna is installed on the roof of the house and is tilted to point to the equatorial plane. The output from the amplifier is passed downstream through about 15~m of RG-58 coaxial cable and fed to the SDR. 




\subsection{Acquisition software}
The SBC runs a server version of Ubuntu 20.04 GNU/linux. Remote access software is used to remotely login and control the system. The acquisition software, which evolved over the course of several months, is based on the popular \texttt{rtl\_power\_fftw} software tool suitably modified to support the bias-T option available in RTL-SDR Blog v3 SDR. This software tool performs FFT on the acquired samples, average the spectra and writes the resulting spectra out. The output data format of \texttt{rtl\_power\_fftw} is a binary file containing the spectra with an associated metadata text file. For long-term archiving and ease of dissemination they are converted to HDF5 format with a custom \texttt{python} code using \texttt{h5py}. The metadata are written to the HDF5 files as attributes, thereby easing further processing. A simple shell script is used to collect spectra over 24 hours using the aforesaid tool, name the file with a human-readable UTC time and perform the \texttt{hdf5} conversion. The script is run with the \texttt{nohup} POSIX command to prevent it from being killed by the OS. The data so collected are downloaded using remote access software for analysis. All further analysis is carried out in \texttt{python}.

\section{Data collection and RFI subtraction}
\label{sec:data_rfi}
The initial spectral scans in various frequency bands showed multiple known emissions such as local and distant FM stations and amateur radio transmissions in the 144-146 MHz. However, also evident in the aforesaid spectra was strong and very frequent RFI. This is not unexpected given the instrument's location in a semi-urban locality; there are multiple power transmission lines nearby, as well as busy roads and electrified railway lines within a few kilometres (for a recent review of urban RFI, see \citeA{7833115}). Analysis of higher time and frequency resolution spectra showed that apart from a few constant frequency tones (perhaps originating from the wired internet connection), RFI is mostly transient in nature with broadband spectra. Such broadband spectra is characteristic of impulsive RFI, such as those from automobile ignition systems. The presence of impulsive RFI precluded sensitive observations in the FM band for meteor scatter, as most meteor reflections typically last for about few milliseconds \cite{1996pimo.conf...99W} and could be easily masked by RFI. Therefore it was decided to pursue ionospheric observations in the $\sim$ 250~MHz satellite downlink band, and explore techniques to mitigate transient RFI with frequency differencing. 

\begin{figure}[!t]
\includegraphics[width=1.0\columnwidth]{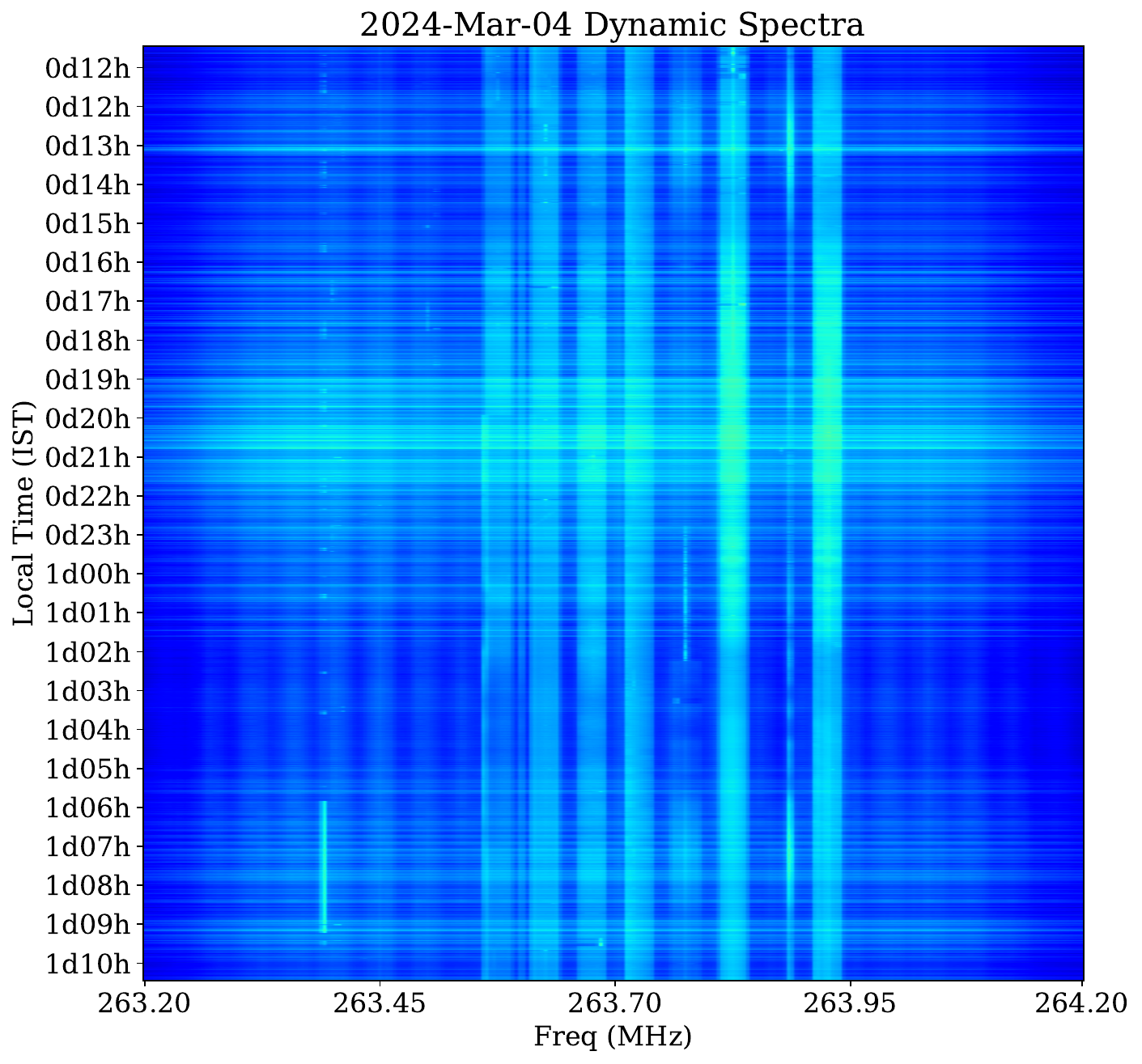}
\caption{Dynamic spectra of the data collected on 2024 March 04. In the y-axis, 0d denotes data collected on the date specified in the title, while 1d denotes data taken after the local midnight, i.e. on the subsequent day.}
\label{fig:waterfall}
\end{figure}

Attempts to collect data for ionospheric studies started in late Dec 2023. The centre frequency was chosen as 263.7~MHz which has strong signals from a geosynchronous satellite UFO-11 located at a longitude of 75.587\degree E. 256 channels over 1~MHz were deemed to be sufficient to measure power in separate narrowband downlink channels, while providing sufficient "blank" channels in between. The data are averaged for 0.5~s. Although 0.5~s is chosen as the averaging time, the actual time between spectra is about 0.6~s due to processing overheads. While higher time cadence could be beneficial for more detailed analysis, averaging for 0.5~s reduced data volume which is beneficial for a remotely operated SBC based system. With these settings, the data volume with 32-bit floating point precision is about 140~MB per 24 hours. 
An example dynamic spectrum ("waterfall") of data collected on 2024 March 04 to 05 is shown in Fig.\ref{fig:waterfall}, in which broadband RFI manifest as horizontal stripes. This specific dataset is chosen for rest of the analysis reported in this paper, as despite the data suffering from severe RFI, the subtraction scheme outlined below removes RFI and reveals scintillations.

\begin{figure}[!t]
\includegraphics[width=1.0\columnwidth]{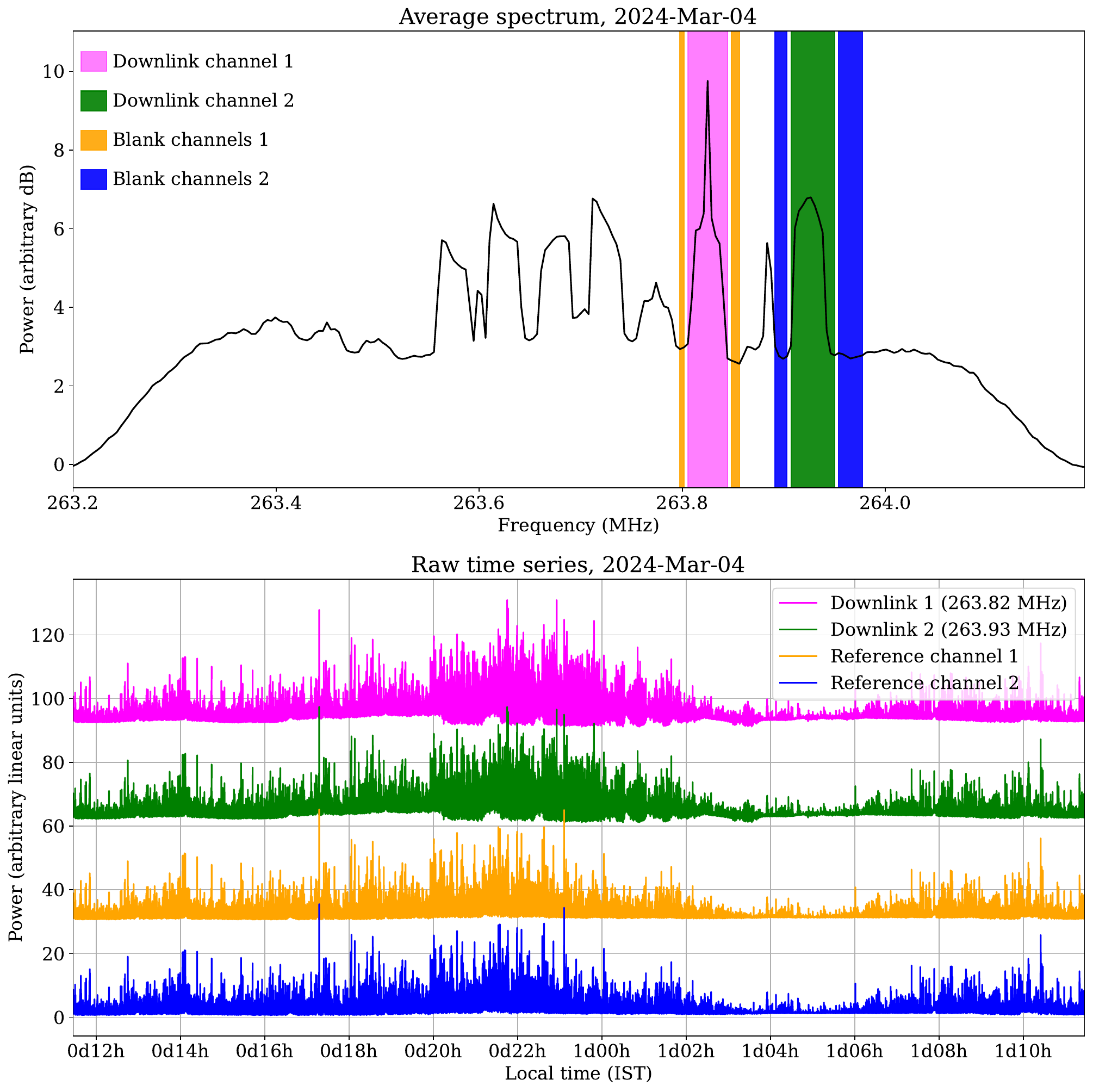}
\caption{In the top panel, the time averaged spectrum of initial 5 minutes of data shown in Fig.\ref{fig:waterfall} is shown. The magenta and green regions are two satellite downlink channels while the orange and blue regions are the adjacent blank channels.
In the bottom panel, raw time series of the two downlink channels and the reference channels (mean of blank channels adjacent to the downlink channels) are plotted, demonstrating the extreme local RFI conditions. The lines are artificially offset for clarity.}
\label{fig:avspec_rfi}
\end{figure}

\begin{figure}[!t]
\includegraphics[width=1.0\columnwidth]{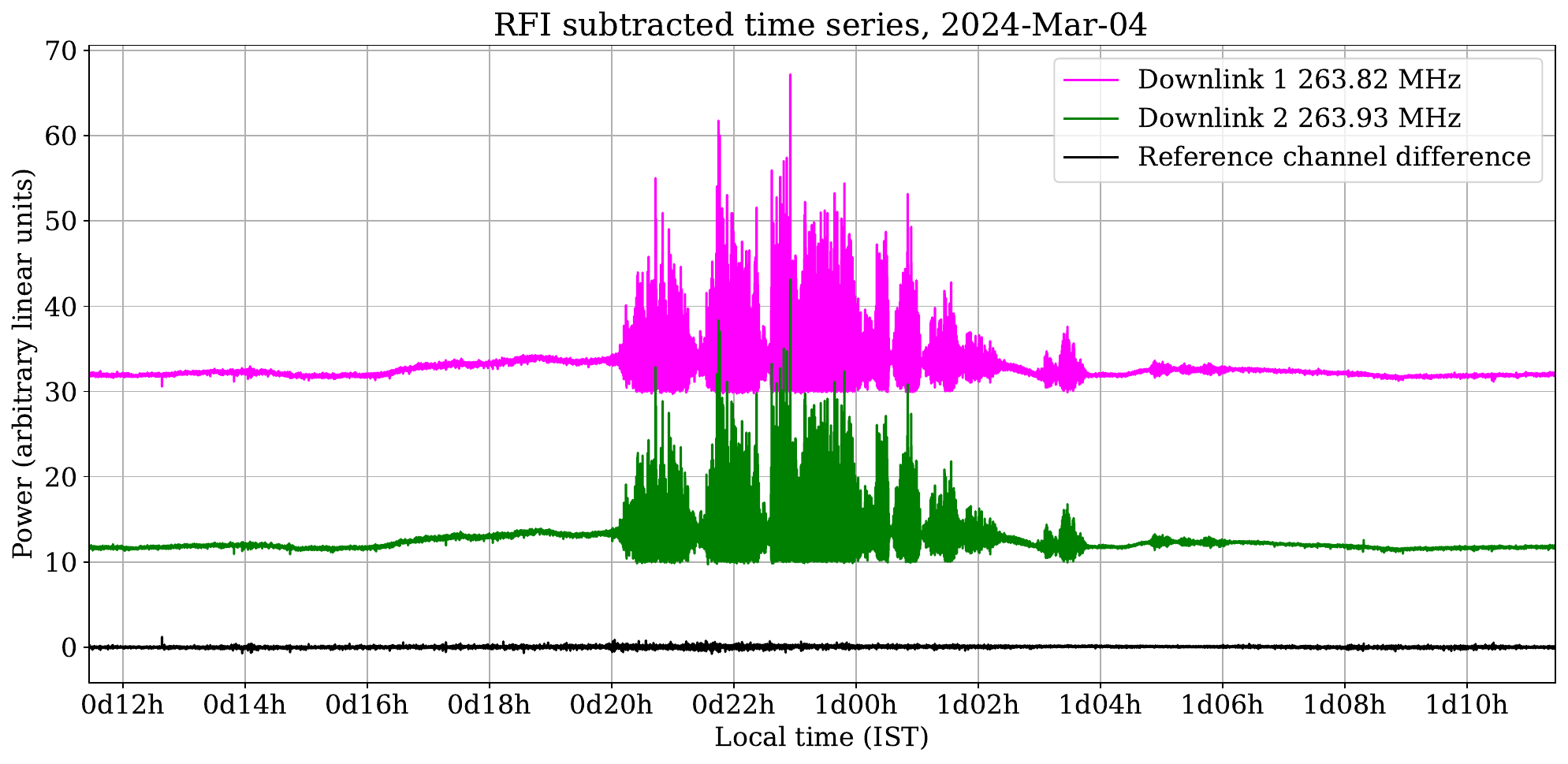}
\caption{RFI subtracted time series from two downlink channels; also shown is the difference between the reference channels used for RFI subtraction. The lines are artificially offset for clarity. Strong scintillations are evident.}
\label{fig:scint_series}
\end{figure}

To further investigate the nature of RFI, two downlink channels and adjacent blank channels, as shown in Fig.\ref{fig:avspec_rfi}, are selected from the data. Their frequency averaged powers as a function of time are also plotted in Fig.\ref{fig:avspec_rfi}. It may be mentioned here that in order to take into account minor variations of the spectral bandpass, for each downlink channel a \textit{reference} channel is formed by taking the mean of the corresponding blank channels on either sides. It is evident from Fig.\ref{fig:avspec_rfi} that the RFI is similar across the different raw time series. This points to the possibility of subtracting the reference channels from the downlink channels - frequency differencing - to remove RFI. This is done in Fig.\ref{fig:scint_series} where it can be seen that the RFI has been eliminated successfully revealing strong scintillations in the received power. In the same figure, the difference between the reference channels is also plotted to demonstrate the effectiveness of RFI subtraction, as well as the lack of scintillations in those channels. 

\section{Results and discussion}
\label{sec:results}
\begin{figure}[!t]
\includegraphics[width=1.0\columnwidth]{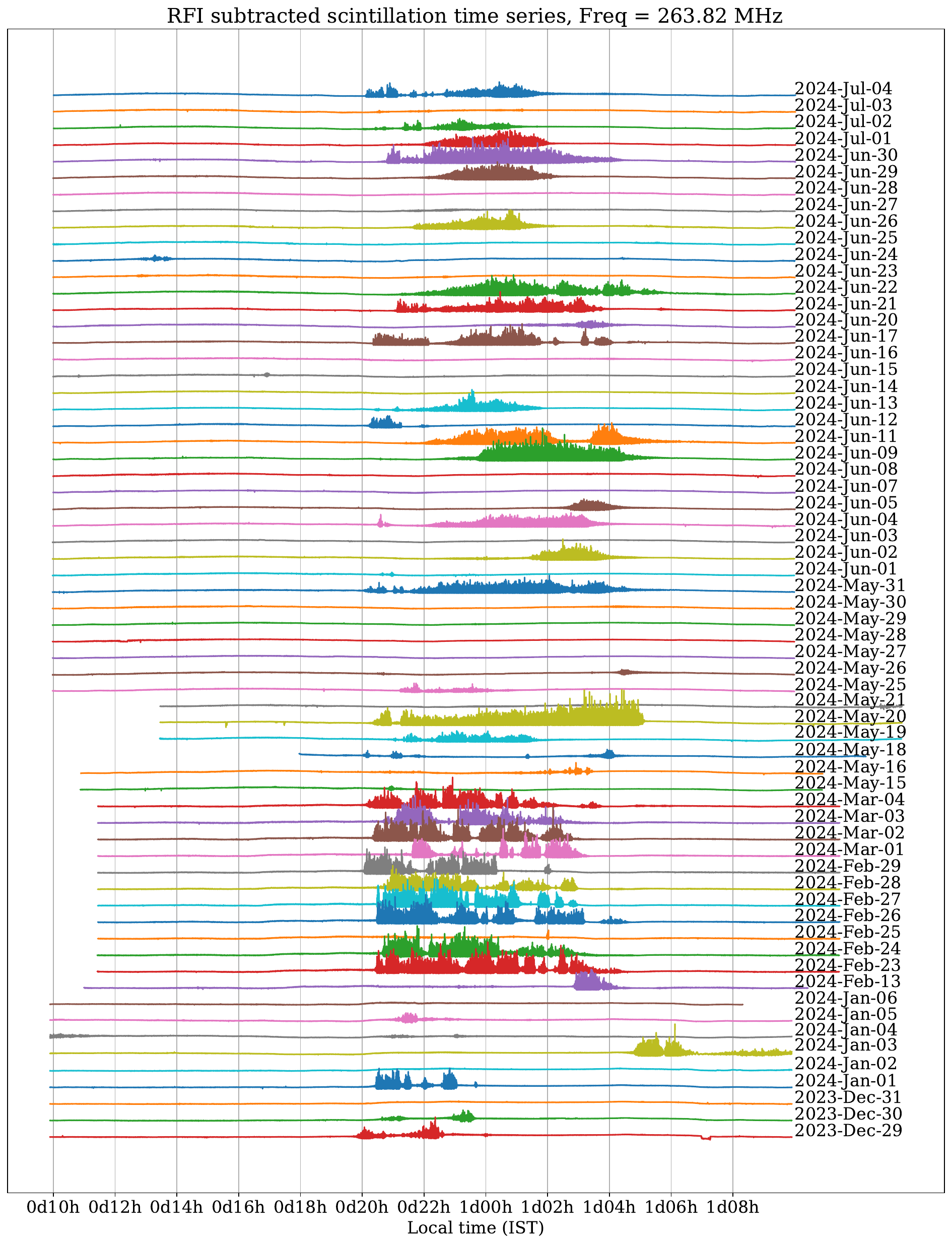}
\caption{Stack plot of RFI subtracted time series across several months. As with previous plots, the lines are artificially offset for clarity.}
\label{fig:offset_plot}
\end{figure}

In Fig.\ref{fig:offset_plot}, a stack plot of scintillations observed in the RFI subtracted data collected till early July 2024 is presented. It is obvious that RFI subtraction performs well most of the time, although on some occasions it leads to minor artefacts. These artefacts can be identified by inspecting the corresponding dynamic spectra, as well as the difference between the reference channels. It is also apparent that the equatorial ionospheric scintillations are predominantly a nighttime phenomena, except for a few weak to moderate daytime scintillations. The scintillations in these data can be broadly classified into two categories \cite{1986JGR....91..270B, 2005AnGeo..23.2457R}. The plasma bubbles formed in the post-sunset hours due to Rayleigh-Taylor instabilities in the ionosphere give rise to scintillations which are called plasma bubble induced (PBI) scintillations. They are characterised by sudden onset in the post sunset hours and appear as several short duration patches. The other class of scintillations are associated with bottomside sinusoidal (BSS) irregularities. These events typically last for several hours as continuous stretches, unlike PBI events which appear as short duration patches. BSS events are also more prominent in the post-midnight hours, while PBI events commence around 20 hrs local time.  

An interesting semi-diurnal pattern became visible after stitching together the data collected over a few days of relatively low ionospheric activity into a continuous time series. As UFO-11 is a geosynchronous satellite, the semi-diurnal pattern is likely to be arising from its orbital motion, with the satellite sweeping through the beam pattern of the receiving antenna. Without advanced techniques, the presence of this semi-diurnal systematic precludes investigations to study diurnal TEC variations or semi-diurnal ionospheric tides.

\subsection{Spectral analysis}
Power spectrum of a scintillation time series is useful in determining properties of the associated plasma irregularities, such as its drift velocity \cite{1972JGR....77.4761R, 1982IEEEP..70..324Y}. The power spectrum of scintillations has some characteristic features which are summarised here. The spectrum peaks at a frequency $f_F$, called the Fresnel frequency which is related to the Fresnel scale  $d_F$ (also known as the Fresnel radius). The spectrum has a $f^{-m}$ shape for temporal frequencies $f>f_F$, which manifests in itself as a straight line in a log-log plot. The power spectrum sometimes shows notches called Fresnel minima at frequencies denoted as $f_1$, $f_2$, $f_3$ .. etc, arising from Fresnel oscillations. These minima are expected to be visible if the vertical thickness of the irregularities is less than 100~km. \citeA{1972JGR....77.4761R} concluded that if the thickness is greater than 50~km smearing would cause less than four minima to appear in the spectrum. The Fresnel scale $d_F$ and the various frequencies are related to the drift velocity of the plasma as follows. 
\begin{eqnarray}
    d_F &= \sqrt{\lambda z}
    \label{eq:fres_scale}
\end{eqnarray}
\begin{equation}
    f_F = \frac{V_0}{\sqrt{\pi \lambda z}}
    \label{eq:f_max}
\end{equation}
\begin{eqnarray}
    f_n &= \frac{V_0\sqrt{n}}{\sqrt{\lambda z}}; n=1,2,3 ..
    \label{eq:fres_min}
\end{eqnarray}
where $\lambda$ is the wavelength, $V_0$ is the horizontal velocity of the irregularities and $z$ is the slant range to the plasma from the observer \cite{1974JATP...36..113S}. Another useful and easy to compute quantity is the scintillation index ($S_4$) which is defined as the ratio of standard deviation of signal intensity to average signal intensity, as given by
\begin{eqnarray}
    S_4 &= \frac{\sigma_I}{\mu_I} = \frac{\sqrt{<I^2> - <I>^2}}{<I>},
    \label{eq:s4_index}
\end{eqnarray}
where $\mu_I$ and $\sigma_I$ are the mean and standard deviation of the signal intensity $I$ computed over 60 second intervals \cite{2019RaSc...54..618B}.

\begin{figure}[!t]
\includegraphics[width=1.0\columnwidth]{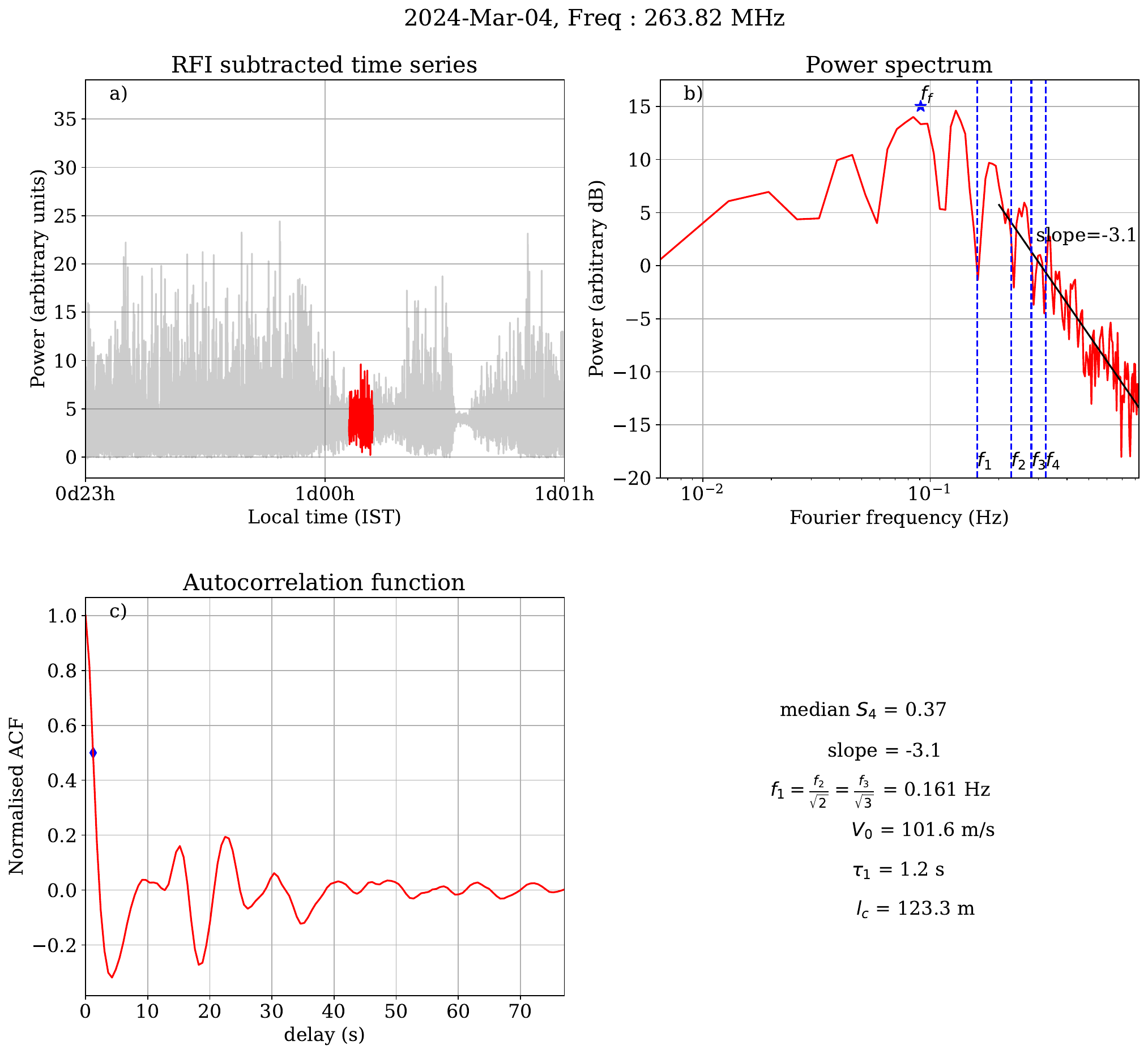}
\caption{Spectral analysis of a 6 minute segment of scintillations. In panel a), the patch used for analysis is highlighted in the time series. In panel b), the power spectrum computed is shown along with the identified Fresnel minima at frequencies $f_1, f_2$, $f_3$ and $f_4$. The Fresnel frequency $f_F$, calculated from $V_0$ and $z$, is also marked. In panel c), the normalised autocorrelation function is provided in which the half decorrelation time is marked. }
\label{fig:scint_fft}
\end{figure}

For spectral analysis, a short segment of 6 minutes around midnight from the data shown in Fig.\ref{fig:scint_series} is used. The median $S_4$ index for this patch is 0.37, implying moderate scintillations. The power spectrum is obtained by calculating a periodogram using Welch's method \cite{1967ITAE...15...70W}. The parameters used for this computation are $N_{FFT}=256$ yielding 129 frequency points, Hann windowing, 50\% overlap between segments, and linear detrending. Results from this spectral analysis are shown in Fig.\ref{fig:scint_fft}.

The different frequencies at which Fresnel minima occur share a well-defined relation of $f_1 = \frac{f_2}{\sqrt{2}} = \frac{f_3}{\sqrt{3}} = \frac{f_4}{\sqrt{4}}$, and can be identified from the power spectrum in Fig.\ref{fig:scint_fft}. Therefore they can be used to estimate $V_0$ using the equation given in Eq.\ref{eq:fres_min}.
Assuming the ionospheric pierce point (IPP) to be 350 km above the receiver, $V_0$ is estimated to be 101.57 $m/s$ for $f_1 = 0.161 Hz$ which is close to the values reported in the literature \cite{2001GeoRL..28..119B, 2005AnGeo..23.2457R}. The estimated value of $V_0$ along with the assumed $z$ is used in Eq.\ref{eq:f_max} to calculate $f_F$ which is marked in Fig.\ref{fig:scint_fft}. It can be seen that $f_F$ is close the maximum in the power spectrum, consistent with the expectations. The IFFT of the power spectrum gives the autocorrelation function (acf) of the timeseries. The delay at which the acf falls less than 50\% of the peak is the half decorrelation time $\tau_1$. With the known drift velocity $V_0$ and $\tau_1$, the characteristic length scale of the irregularities can be determined as 123.3~m. This corresponds to intermediate scale irregularities \cite{2006JASTP..68.1116S}. The presence of Fresnel minima also shows that the irregularities are confined to a layer approximately 50~km in thickness. It is interesting to note that the presence of such Fresnel minima is often associated with BSS irregularities. Therefore although the data may resemble PBI scintillations in the immediate post-sunset hours, the scintillations in the post-midnight hours are probably due to BSS irregularities. Similar occurrences of BSS events following PBI events have been observed in the past - see for e.g. \citeA{1986JGR....91..270B}. Alternatively, these irregularities may be thin ($\sim$ 50~km), but bubble induced.

\subsection{Daytime scintillations}
\begin{figure}[!t]
\includegraphics[width=1.0\columnwidth]{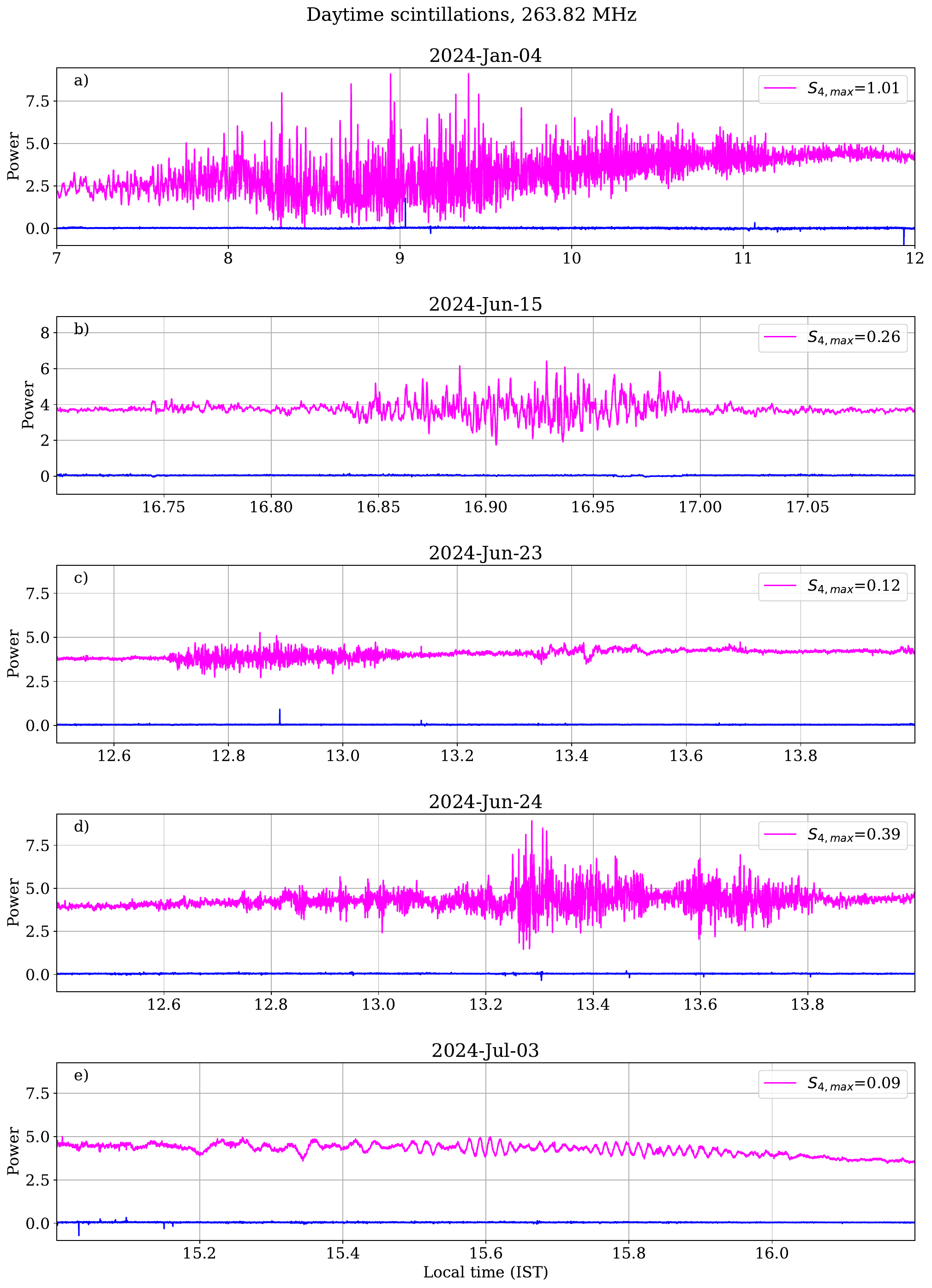}
\caption{A collection of daytime scintillation events along with their peak $S_4$ values. In panel a), a strong post-sunrise scintillation event that lasted for about 5 hours is seen. In panels b), c) and d), weak to moderate post-noon scintillations around the summer solstice in the northern hemisphere are seen. In panel e), the weak scintillations are quasi-periodic (QP) in nature. In all panels, the difference between reference channels is also plotted to check the quality of RFI subtraction.}
\label{fig:dt_scint_collage}
\end{figure}

Daytime scintillations near the magnetic equator are related to the irregularities in the E-layer such as sporadic E-layers and blanketing sporadic E-layers ($E_{sb}$) \cite{1989IJRSP..18..121K}. Some of these events are also associated with eastward flowing counter equatorial electrojet (CEJ) during daytime \cite{1980IJRSP...9..219R}. Daytime equatorial scintillations are somewhat rare; for example \citeA{2015JGRA..120.9074Y} analysed scintillation data from 2000-2006 collected at Tirunelveli, India and found only seven events - all in the post-noon hours during June–July solstice of 2003–2005. Therefore collection of such daytime scintillation data is valuable in understanding the complex dynamics of the equatorial ionosphere and the associated current systems. In Fig.\ref{fig:dt_scint_collage}, a collection of a few daytime scintillation events seen in the PIE data is presented. In this paper outlining the initial results, only the time series and corresponding maximum $S_4$ are provided; detailed analysis of these events is left for future works.


%

\section{Conclusions}
In this paper, a low cost system to monitor the equatorial ionosphere is presented. Despite the presence of strong RFI, initial results from the system reveal ionospheric scintillation. The results show that the system is also sensitive to weak daytime scintillations of various types. Owing to the use of a SDR as the receiver, inexpensive systems similar to the one described in this paper can be easily deployed by universities, educational institutions, radio amateurs and citizen scientists. 
Most the design elements of this experiment are non-critical, however the data capture software requires a USB dongle based on RTL2832U. Antennas with more gain such as multi-element Yagi-Uda antennas can be used. Instead of Orange Pi Zero2 other SBCs like Raspberry Pis can be used. The pre-amplifier can be any low noise amplifier with unconditional stability. 

This work is similar to other low-cost ionospheric experiments such as the ScintPi \cite{2019E&SS....6.1547R, 2022EP&S...74..185G}. However, unlike the ScintPi which is an L-band GNSS experiment, the experiment described in this paper is in the $\sim$ 250~MHz geosynchronous satellite downlink band which is more sensitive to ionospheric conditions than L-band. On the other hand due to their use of multiple GNSS satellites L-band experiments can probe several lines of sight simultaneously, while the geosynchronous satellites are fewer in number and provide limited lines of sight. Therefore, combining the experiment proposed in this paper with other experiments such as the ScintPi can be an extremely powerful citizen science project, resulting in a crowd-sourced data repository like the Personal Space Weather Station (PSWS) network \cite{2023ESSD...15.1403C}. 

The resulting dataset can provide a wealth of information on the equatorial and low-latitude ionospheric conditions. For example, by performing cross-correlation of identically acquired data  collected with multiple instruments spread over a large region, independent estimates of the drift velocity can be acquired. Other secondary uses of such a spectral dataset could be understanding the nature of RFI conditions, utilising the blank channels in the data. A cursory inspection of the PIE data shows that the RFI has distinct patterns such as high activity during daytime and much lower occurrences in the early mornings. In conclusion, it is also expected that enthusiastic citizen scientists replicate the results reported in this paper.  



\begin{acronyms}
    \acro{RFI} Radio frequency interference
    \acro{CEJ} Counter equatorial electrojet
    \acro{PBI} Plasma bubble induced
    \acro{BSS} Bottomside sinusoidal irregularities
    \acro{FFT} Fast Fourier transform
    \acro{IFFT} Inverse fast Fourier transform
    \acro{GNSS} Global navigation satellite system
    \acro{SBC} Single board computer
    \acro{SDR} Software defined radio
    \acro{TEC} Total electon content
    \acro{IST} Indian standard time
\end{acronyms}

\section{Open Research}
The raw data used for this paper are available in a dataset associated with this publication on Zenodo \cite{zenodo_upload}. The current plan is to periodically update the the Zenodo repository with data collected from the PIE. Codes and scripts used in this paper are available at \url{https://github.com/jnthek/pie-codes/}. The data acquisition software is based on \texttt{rtl\_power\_fftw}, the original version of which is available at \url{https://github.com/AD-Vega/rtl-power-fftw} and the bias-T modified version used for this experiment is available at \url{https://github.com/jnthek/rtl-power-fftw}.


\section*{Acknowledgement}
The author would like to acknowledge the unwavering support provided by his uncle Mr. Vishnudas T. E. during the construction and operation of the experiment described in this paper. PIE was constructed by the author using personal funds, however the highly stimulating academic environment provided by ICRAR/Curtin is gratefully acknowledged. The author's colleagues at ICRAR/Curtin provided encouragement and useful discussions. This work makes use of the NASA Astrophysics Data System (ADS). This work also uses the following python packages and the authors and maintainers of these packages are thanked - \texttt{numpy }\cite{2020Natur.585..357H}, \texttt{scipy} \cite{2020SciPy-NMeth}, \texttt{matplotlib}\cite{Hunter:2007}, and \texttt{h5py} (\url{https://www.h5py.org/}).

------------------------------------------------------------------------ 

\bibliography{ionosphere}

%

\end{document}